\documentclass[12pt,authoryear]{elsarticle}

\usepackage{graphicx}  
\usepackage{amssymb}
\usepackage{amsmath}
\def\d{\mathrm{d}}
\def\xR{\mathbb{R}}
\def\x0{\mathbf{0}}

\def\L{\mathcal{L}}

\usepackage{xcolor}

\begin{document}

\title{Quasi-static shape control of soft, morphing structures}

\author{Eszter Fehér\fnref{fn1,fn2}}
\ead{feher.eszter@epk.bme.hu}
\author{András Á. Sipos\fnref{fn1,fn2}}
\ead{siposa@eik.bme.hu}
\author{Péter L. Várkonyi\fnref{fn3} \corref{cor1}}
\ead{varkonyi.peter@epk.bme.hu}

\fntext[fn1]{Department of Morphology and Geometric Modeling, Budapest University of Technology and Economics, Muegyetem rkp. 3, H-1111 Budapest, Hungary}
\fntext[fn2]{HUN-REN-BME Morphodynamics Research Group, Budapest University of Technology and Economics, Muegyetem rkp. 3, H-1111 Budapest, Hungary}
\fntext[fn3]{Department of Mechanics, Materials, and Structures, Budapest University of Technology and Economics, Muegyetem rkp. 3, H-1111 Budapest, Hungary}
\cortext[cor1]{Corresponding author}

\begin{abstract}
Inspired by biological systems, we introduce a general framework for quasi-static shape control of human-scale structures under slowly varying external actions or requirements.  In this setting, shape control aims to traverse the stable sub-manifolds of the equilibrium set to meet some predefined requirements or optimization criteria. As finite deformations are allowed, the equilibrium set may have a non-trivial topology. 
This paper explores the implications of large shape changes and high compliance, such as the emergence of unstable equilibria and equilibrium sets with non-trivial topology. We identify various adaptivity scenarios, ranging from inverse kinematics to optimization and path planning problems, and discuss the role of time-dependent loads and requirements. The applicability of the proposed concepts is demonstrated through the example of a curved Kirchhoff rod that is susceptible to snap-through behavior.
\end{abstract}

%
%

\maketitle


\section{Introduction}

In principle, human-designed civil engineering structures withstand external actions by their robustness, i.e., without significant shape changes. It is required that the anticipated variations in external actions do not lead to structural failure. Adaptation, i.e., favorable modifications of the geometry or the material performance of the operating structure, is not considered in classical design methodologies.

Disregarding the practical challenges posed by such variability, structural adaptability offers significant advantages: unanticipated types of actions or unexpected magnitudes are less likely to precipitate structural failure. In other words, the resistance of the structure is not determined per se at the time of execution, leaving room to tune the risk of failure during the operation. 
Adaptive structures implement this concept by adjusting their shapes or mechanical properties in response to changing loads or requirements during service, utilizing sensors, actuators, and control laws. The typical objectives of adaptation are to reduce stresses \citep{shen2013static,shen2022static}, deformations \citep{steffen2022actuation}, or vibrations, or to achieve shape-related targets \citep{sachse2021variational}. The application of adaptive systems has a long tradition in earthquake protection of high-rise buildings \citep{morales2013towards}. However, other applications are still in an early phase. Notable examples of experimental projects include active compensation of static deformations of slender beams \citep{senatore2018shape,burghardt2022investigation} and slabs \citep{nitzlader2024experimental}; vibration suppression of thin shells \citep{weickgenannt2012active,neuhaeuser2013stuttgart}, slender beams \citep{del2023design} and high-rise buildings \citep{blandini2022d1244}; as well as optimization of stress fields either by adaptive prestress \citep{senatore2018exploring} or by shape morphing, in the case of trusses \citep{reksowardojo2019experimental,reksowardojo2020design} and arches \citep{van2016adaptive}.  In many cases, adaptation results in a significant reduction of whole-life energy consumption \citep{reksowardojo2023design,senatore2018exploring,senatore2019synthesis}. 

The leading inspiration to build adaptive structures without doubt comes from \emph{biological systems}. On the one hand, adaptation is central to evolution; however, it occurs slowly through hundreds of generations. In the engineering sciences, adaptation in evolution might be associated with the development of historical structures through the ages \citep{heyman1995}.
Adaptiveness at a short time scale also appears in nature. Think, for example, of the environmentally sensitive growth of plants and animals \citep{goriely2017mathematics,oliveri2024}. The stress-dependent development of skeletons and the bone microstructure in specific \citep{currey1970} are analogous to adaptive material properties in engineering systems. The tropic behaviour of plant roots and shoots, like sunflower turning towards the Sun, the directional growth of shoots against gravity, and the development of intrinsic curvature in response to touch in the case of climbing plants \citep{oliveri2024}, all show analogy with engineering structures subject to shape adaptation.  

The modeling of adaptive structures relies heavily on \emph{continuum mechanics}, addressing the deformations of solids under external loading \citep{timoshenko1951,antman1995}. Adaptiveness in the realm of continuum mechanics covers all cases when the body's reference (unloaded) configuration changes over time via some control mechanism. 

Quasi-static systems subject to sufficiently slow changes of loads and shape evolve along equilibrium states, which form the \emph{equilibrium set}.
 Structures that undergo large shape changes either due to softness or as a result of adaptation require \emph{geometrically exact} models to faithfully represent the interaction between their deformed geometry and external loading \citep{antman1995}. The interplay between the deformed geometry and the internal stresses results in nonlinear models, where the equilibrium set might be highly complicated with bifurcating branches and non-uniqueness of the solutions \citep{guckenheimer1983,zeidler1985}.  

Although establishing the equilibrium set is computationally challenging \citep{domokos1995, domokos2004}, it carries essential information about the regions of stable solutions and sensitivity to imperfections. Specifically, the computation of stable equilibrium solutions has been at the forefront of research for decades \citep{pogorelov1988, kumar2010}. Avoiding \emph{snap-through}, either in the discrete or the continuous setting, has a distinguished role in these studies. Systems prone to snap-trough addresses the question of multi-stability \citep{chai2024} and the sudden emergence of dynamical response under slow variation of the parameters \citep{liu2021}.
In the case of the elastica (a slender, unshearable, inextensible elastic rod prone to large deformations), in the absence of lateral loading and a planar setting, stability maps can be obtained analytically \citep{cazzolli2019,cazzolli2020}. In some special cases, snap-trough is purposefully designed to reach specific goals, for example, in multiphysics problems \citep{amor2025,tan2025}, the design of metamaterials \citep{azulay2023}, or reconfigurable morphing structures \citep{rahman2024}. 

%

\begin{figure}[ht!]
    \centering
    \includegraphics[width=0.7\linewidth]{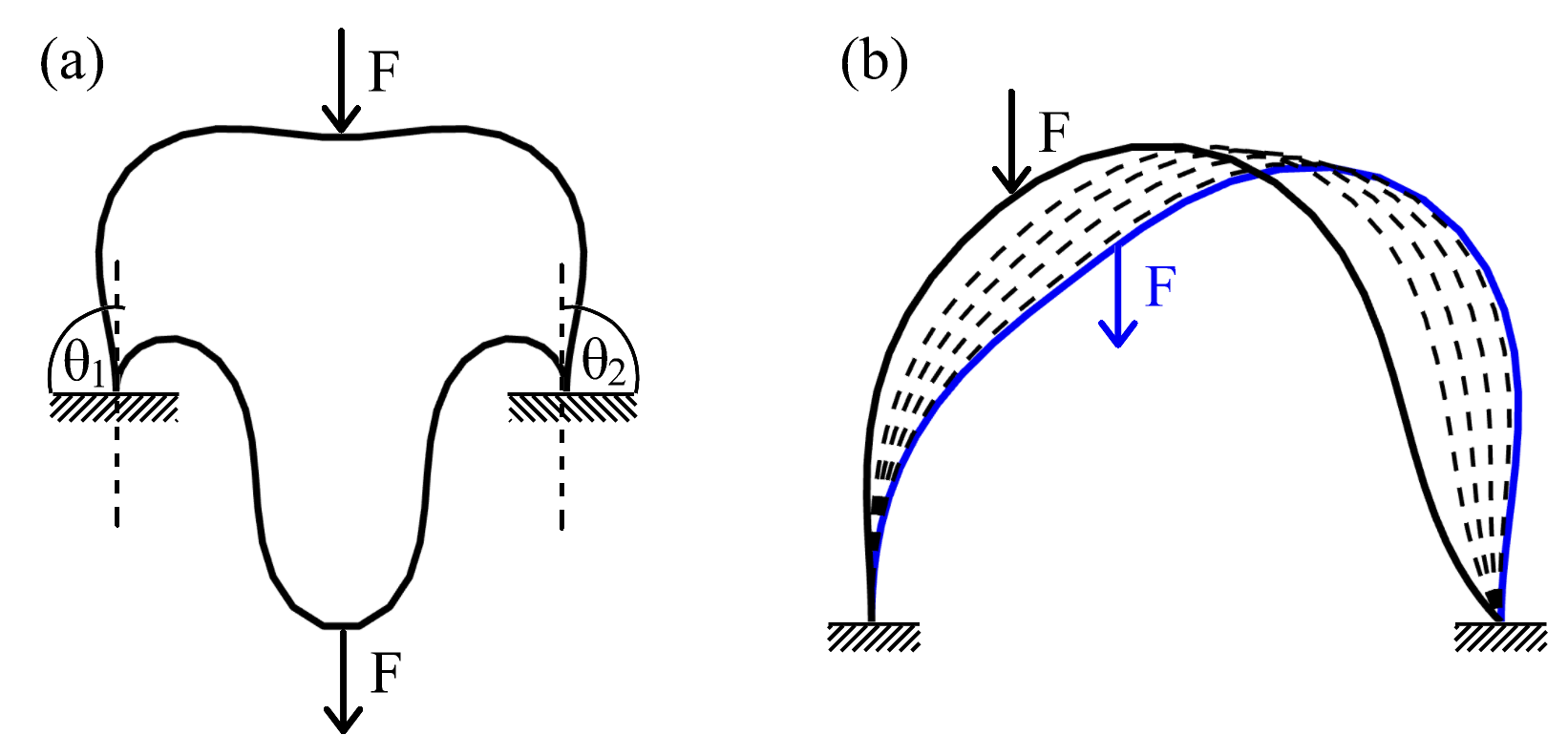}
	\caption{a) A clamped-clamped curved rod with prescribed end point tangents $\theta_1$ and $\theta_2$ carrying a concentrated load $F$. The system has two stable configurations for the same set of parameters $F, \theta_1, \theta_2$, indicating non-uniqueness in the parameter space. b) Shape adaptation of the curved rod. The initial configuration is marked in blue. After the load is applied, the $\theta_1, \theta_2$ angles at the supports change continuously to reach the target configuration. Dashed lines mark the in-between states, and the thick black line marks the target state.Quasi-static shape adaptation aims to find a path between the initial and target shapes such that all intermediate configurations are stable.}
	\label{fig:intro_example}
\end{figure}

In this paper, we focus on quasi-static adaptive structures. We go beyond existing works by allowing that the structure is soft, making it susceptible to instability and snapping (Fig. \ref{fig:intro_example}a). Moreover, we allow for large shape changes, giving rise to strong nonlinearity and equilibrium sets with non-trivial topology. Within this framework, we introduce a classification of tasks ranging from standalone \emph{shape optimization tasks} and \emph{inverse kinematics problems} to parametric families of such tasks (due to time-dependent load or requirements) as well as \emph{path planning} problems. We present a technique to solve shape adaptation tasks by computing the equilibrium set and then finding stable paths connecting points corresponding to different states of the structure (Fig. \ref{fig:intro_example}b). 
It is highlighted that working with the equilibrium sets of complex problems is a challenge since it is possible that multiple possible stable states correspond to the same control parameters, which poses non-uniqueness in the parameter space (Fig. \ref{fig:intro_example}a). We propose multiple methods to resolve this issue.
We note that the idea of using the equilibrium set in shape optimization of rod problems has already been used in the robotics community \citep{bretl2014}. 



The presented techniques and tasks are illustrated by simple examples, including a single degree of freedom discrete elastic system, as well as a continuous curved rod model representing a \emph{soft arch}. 
With a strong motivation from engineering problems, we study cases when control of the shape is restricted to a finite number of actuators. On the one hand, this provides a feasible setting for human-made structures; on the other hand, a finite number of parameters need to be controlled. Our rod problem is similar to that of Bommel et al. \citep{bommel2016}, who calculated the optimal support angles of an arch made of a flat sheet under concentrated load by combining the finite element method with simulated annealing. Nevertheless, we allow larger deformations, triggering snap through and the associated challenges. For this example, the proposed  workflow is used to solve various tasks, including a smooth transition between the upper and inverted lower position, avoiding snap-through
(Fig. \ref{fig:intro_example}a) 
by tactically and continuously varying the support angles. 
In Sec. \ref{sec:methodology}, the essence of the proposed technique, including an illustration problem, is presented. Multiple optimization tasks are solved in Sec. \ref{sec:application}. Finally, in Sec. \ref{sec:conclusion} we conclude and present some insights about the future smart structures.

\section{Methodology} \label{sec:methodology}
\subsection{The equilibrium set} \label{sec:equilibrium}

The \emph{equilibrium set} is a set of points (more precisely, a manifold), each point of which corresponds to a configuration of the structure together with a specific set of loads, such that the structure is in static equilibrium. In the case of quasi-static shape adaptation, the equilibrium set represents all possible equilibrium states of the system, and shape adaptation can be understood as navigation along the equilibrium set.

\begin{figure}[ht!]
\centering 
    \includegraphics[width=\textwidth, angle=0]{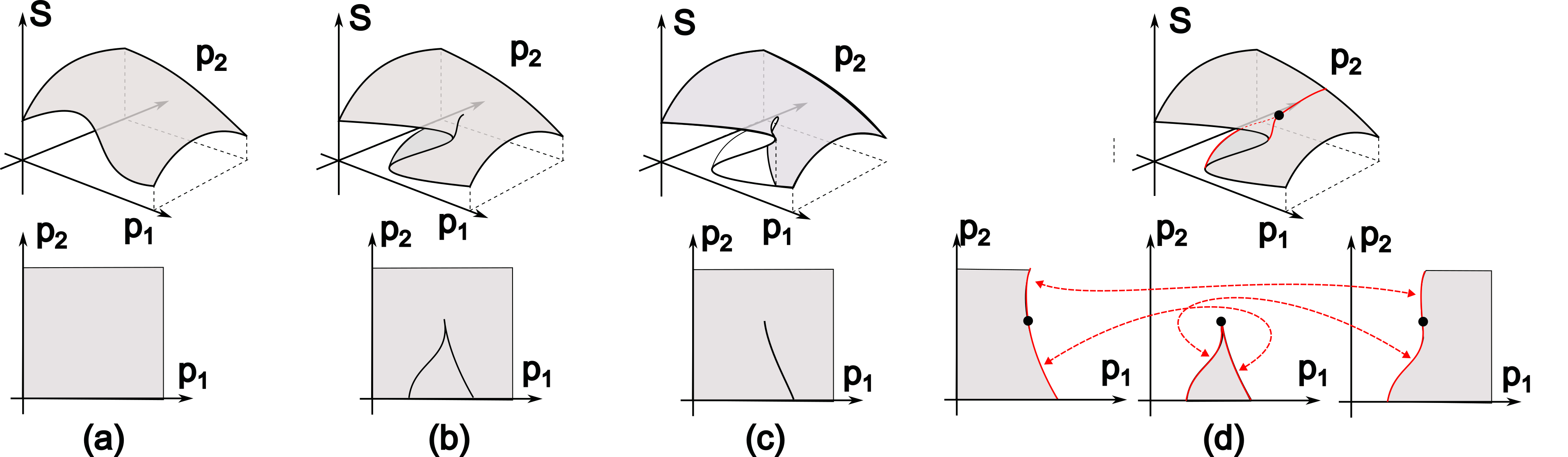}	\\
	\caption{Examples of equilibrium sets (top) and their internal parameterizations (bottom). Each panel in the top row shows an equilibrium surface embedded in 3D space spanned by two actuator and/or load parameters ($p_1,p_2$) and a shape descriptor ($S$). (a) internal parameterization by projection into the subspace of parameters. (b) projection fails if multiple states correspond to the same parameter values (i.e., non-uniqueness). (c) restoring uniqueness by reduction of the equilibrium set to the shaded part of the surface. The thick curve at the bottom is a barrier curve.  (d) restoring uniqueness by dissection of the equilibrium set to 3 components with distinct internal parameterizations for each one.} 
	\label{fig:param}
\end{figure}

The dimension $d$ of the equilibrium set is typically
\begin{align}
\label{eq:d:def}
d=d_a+d_l
\end{align}
where $d_a$ is the total number of \emph{actuators} and $d_l$ is the number of \emph{load parameters}. The two types of parameters play different roles, as actuator parameters are typically tuned during the adaptation process, whereas load parameters are determined independently by environmental conditions (and are often time-dependent). Note that $d$ is independent of the $d_f$ degrees of freedom of the structure; however, $d_a\leq d_f$ except in special situations, when the actuator setup is redundant, and $d_a=d_f$ is needed for full controllability of the shape of the structure \citep{varkonyi2025shape}.
A discrete approximation of the equilibrium set can be generated and stored as a simplicial grid, such as a triangulated surface in the case of $d=2$. For each node of the grid, one should store sufficient information to reconstruct the corresponding structural configuration. Once the equilibrium set has been determined, many shape adaptation tasks can be solved by interpolation. Visualization of the equilibrium set can be enhanced by using an embedding into $\mathbb{R}^{d+d_S}$ spanned by the $d$ parameters mentioned above and $d_S\geq 1$ additional \emph{scalar shape descriptors} (Fig. \ref{fig:param}A). Here, shape descriptor refers to any scalar function that depends on the structure's shape.
  
Because of the geometrically exact mechanical model and possibly the nonlinear constitutive law, the geometry and topology of the manifold associated with the equilibrium set are not simple: self-intersections, disconnected submanifolds, and singularities cannot be excluded, i.e., the equilibrium set is not isomorphic to any subset of the Euclidean space. Note that actuators may induce large motion or rotation, an additional cause of geometric nonlinearity that may emerge even under small deformations and strains. Most importantly, the coexistence of multiple equilibria for the same load and actuator setting, called non-uniqueness in the sequel, occurs. Still, because of continuity, the subsets of the equilibrium set are isomorphic to subsets of $\mathbb{R}^d$, enabling one to traverse curves on the manifold. Hence, solving some more advanced tasks related to the equilibrium set relies on a parameterization of its points, ideally with a non-redundant set of $d$ scalar parameters \citep{sachse2021variational}. 
In the sequel, this will be referred to as \emph{internal parameterization}.

In many simple cases, the internal parameterization of the equilibrium set can be done by using the the actuator and load parameters (Fig. \ref{fig:param}(a)), whereas this approach may fail in other cases due to non-uniqueness (Fig. \ref{fig:param}(b)).

 There are several possible techniques to eliminate non-uniqueness. In some cases an appropriate mapping of the equilibrium set into $\mathbb{R}^d$ does the job. For example, a valid internal parameterization would be obtained in the case of Fig. \ref{fig:param}(b) by projecting the equilibrium set into the plane spanned by $(S,p_2)$ instead of $(p_1,p_2)$.
 
 Another possible approach is \emph{reduction} of the equilibrium set, which means that some parts of it are discarded. 
 For example, one can choose for each actuator setting one preferred equilibrium state (Fig. \ref{fig:param}(c)). 
%
Reduction techniques may introduce special features into the equilibrium set 
 such as the emergence of \emph{barriers} in the projected equilibrium set, where nearby points correspond to distinct points of the original manifold (see Fig. \ref{fig:param}(c) for an example). Such barriers should be taken into account in all tasks where a continuous path between two points is generated. Reduction techniques may also result in the emergence of \emph{gaps}, i.e., ranges of the parameters, which do not correspond to a point of the equilibrium set.

 Instead of reduction, a complex equilibrium set can also be parameterized internally by dissecting it into multiple parts, similarly to the application of \emph{atlases} in classical differential geometry \citep{lovett2019}, with a local internal parameterization for each part, and one-to-one maps defining the corresponding pairs of parameter values at the individual cuts (Fig. \ref{fig:param}(d)).







%

\subsection{Classification of shape adaptation tasks}

Traditionally, a shape adaptation task consists of finding point(s) within the equilibrium set that match a particular goal. Hence, in what follows, we will refer to points of the equilibrium set, as well as the associated internal parameter values, as \emph{solutions} of the shape adaptation task. In the case of problems with a complex equilibrium set, the path between the initial configuration and the solution is non-trivial. Hence, solving the shape adaptation task requires finding this path as well. These observations inspire the classification outlined in the sequel.

\subsubsection{Inverse kinematics and optimization tasks}

Some shape adaptation tasks admit perfect solutions. For example, one can prescribe the values of $d_a$ shape descriptors, in which case a perfect match is often possible by tuning $d_a$  actuators. Such tasks can be formulated as \emph{inverse kinematics} problems, which boil down to root-finding of nonlinear systems of equations. For example, in Section \ref{sec:inversekinematics} we will consider a soft elastica beam,  where the Cartesian coordinates of the centerline are $(x(s),y(s))$, and $s$ denotes arclength along the beam. We prescribe the heights of two specific points, i.e., $y(s_1)$, $y(s_2)$, and find perfect solutions by setting two rotary actuators at the terminal points. However, nonlinearity implies that the solution may or may not exist, and it can be non-unique as well.

In other cases, a perfect solution is not expected or is even undefined; however, the quality of a solution can be quantified by an \emph{objective function} assigning scalar values to the equilibrium states, giving rise to a standard \emph{optimization problem}. For example, this situation occurs naturally if the number of actuators is lower than the number of prescribed shape descriptors. 

 The objective functions typically reflect requirements related to the deformed shapes and internal stress fields of the structure, which often stem from practical safety and serviceability requirements. 
 For example, in Section \ref{sec:mincurve}  we will use the largest curvature along an elastica beam
\begin{eqnarray}
   O_1=\max_s|\kappa(s)|.  
   \label{eq:O1kappa}
\end{eqnarray}
as objective. $O_1$ is - in the case of a straight natural shape - proportional to the maximum stress induced by bending. 
We will also consider another objective for a problem involving a load parameter $Q$ (Section \ref{sec:maxload}), defined as the largest value of $Q$ for which the structure remains in stable equilibrium such that no snap-through occurs.

\subsubsection{Constrained problems}
Feasibility constraints may optionally be involved in inverse kinematics as well as in optimization problems. In the first case, they simply partition potential solutions to feasible and infeasible ones, whereas constraints in optimization problems can be addressed by standard techniques of constrained optimization \citep{pedregal2004introduction}.

 For example, the \emph{gaps} 
 introduced during the internal parameterization of the equilibrium manifold can be handled as feasibility constraints. In addition, there are also physically inspired feasibility constraints in many problems. As the equilibrium sets include stable and unstable ranges, the requirement of \emph{stability} is a common feasibility constraint. 
If a structure reaches a state subject to elastic instability, a dynamic response, specifically \emph{snap-through}, is triggered, which drives it to a distant stable equilibrium state. The dynamic response is often dangerous, and it cannot be modeled using the quasi-static approach of this paper.  
The stability constraint will be included in each example discussed in the sequel. Other examples of feasibility constraints include threshold values of the objective functions mentioned in the previous point, such as a prescribed maximum of curvature in the case of a slender elastica rod.

\subsubsection{Parametric problems} 
\label{sec:parametric}

Slowly and continuously varying time-dependent targets as well as external loads can be represented by a parametric family of inverse kinematics or optimization problems, each of which seeks the desired state of the structure at a specific instant. Even though each of those problems can be solved independently, parametric problems have a crucial additional requirement: their solutions should depend continuously on the parameter in order that the system can track the time-dependent target. 

As an example, consider an equilibrium manifold $\mathbb{E}$ internally parameterized by $d$ internal parameters $p=(p_1,p_2,...,p_{d})$ 
as well as
a set of sufficiently smooth shape descriptor function collected in vector $S(p)$. 
We aim to solve a parametric inverse kinematics problem with time $t$ ($0\leq t\leq T$) as a parameter, where the structure must follow states with sufficiently smooth prescribed values $S^*(t)$ of the shape descriptors: 
\begin{align}
  S(p(t))=S^*(t).   \label{eq:param_inv_kin}
\end{align} 
In order to solve this problem, one needs first a solution $p_0$ to the inverse kinematic problem for time $t=0$. Then, the full parametric problem can be treated as an initial value problem. The governing equation is obtained by differentiation of \eqref{eq:param_inv_kin}, yielding
\begin{align}
\dot p = J_S^{-1} \dot s.
\end{align}
Here, 
 dot means differentiation with respect to time, and $J_S$ is the Jacobian matrix of $S$ with respect to its variables $p$. Clearly, the existence and uniqueness of the solution are guaranteed if $J_S$ is nonsingular. The condition of singularity 
 \begin{align}
      \det(J_S)=0 \label{eq:detJ}
 \end{align}
is typically satisfied along a co-dimension 1 manifold of the equilibrium set, which divides it into distinct regions $E_1, E_2,...$. If the target function $S^*(t)$ is contained entirely by the image of one of the individual regions $E_1, E_2,...$ under the map $S(p)$, then the parametric problem has exactly one solution with $p(0)=p_0$. In the converse case, it is possible that the continuously changing target $S^*(t)$ cannot be reached by a continuous adaptation $p(t)$ of the structure. This is why for any point $e$ in the equilibrium set, the image of a component $E_i$ containing $e$ under the map $S$ will be referred to as the \emph{reachable set} associated with point $e$. An illustrative example will be given in Sec.~\ref{sec:inversekinematics}.

In the case of a parametric optimization problem, the continuity of the optimum may break down due to several reasons, including bifurcations of local optimum values on the equilibrium set.

\subsubsection{Path planning problems} 
In contrast to parametric problems, a path planning problem consists of two prescribed terminal states, and the task is finding a continuous connection between those points along the equilibrium manifold, possibly subject to feasibility constraints. Typically, infinitely many connections exist, and thus, the path may also be subject to optimization. Clearly, the existence of a viable path depends on the connectivity of the equilibrium manifold (or the subset of that manifold consistent with the feasibility constraints), and the complexity of the search algorithms may increase significantly if the equilibrium manifold has a complex topology. Path planning between points belonging to disconnected components of the equilibrium manifold is not possible unless snapping of the structure from an unstable equilibrium to a distant stable one is allowed.    Path planning (either in physical space or in an abstract configuration space) is a common task of robotic manipulators and vehicles; however, it usually happens on a simple manifold with a fixed topology \citep{murray2017mathematical}.

Path planning tasks may have various applications in the context of adaptive structures. For example, imagine that a structure has two distinct modes of operation, which require different shapes. The two modes may correspond to normal use and maintenance, or normal use and a safety mode, which eliminates the risk of failure in the case of an unfavorable loading scenario. 



\subsubsection{Uncertainty}

An important feature of all tasks related to adaptive systems is the type of information available about the state of the system. For structures intended to carry static loads in an efficient way, it is a crucial question whether the external load function is known or not. In the case of multi-stability, one may also seek information about the current shape of the structure. The examples introduced below do not address the question of how load can be measured directly or inferred from indirect measurements. Instead, the load as well as the current state of the structure are treated as known. A related work in \cite{guerra2025} addresses shape optimization in those situations where the load is not directly measurable, but it has an explicit effect on the target to be reached.

\subsection{Illustrative example: a constrained spring}
\label{subsec:simple:example}
\begin{figure}[ht!]
    \centering
    \includegraphics[width=0.30\linewidth]{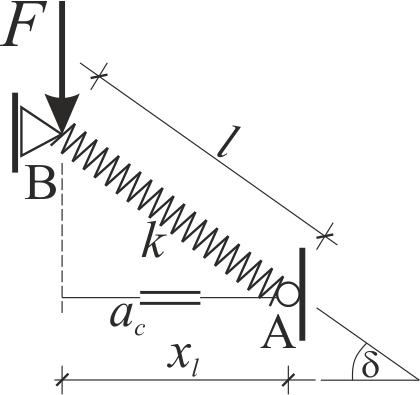}
    \caption{Discrete illustrative problem: A spring with stiffness $k$ is attached to point $A$. Its other end at point $B$ can move along a vertical line at a distance of $x_l$ from $A$. The $a_c$ actuator controls $x_l$ while point $B$, supported with a roller, is loaded with a vertical load $F$.}
    \label{fig:dmodel}
\end{figure}

As an illustrative example, we first present a simple system, as depicted in Figure~\ref{fig:dmodel}, where the equilibrium set can be found analytically. At the same time, this example features a non-trivial equilibrium set that exhibits instability and snap-through. We assume the spring is linearly elastic with stiffness $k$, and its stress-free length equals $1$. The position of endpoint A is fixed at point $(x_l,0)$ where $x_l$ is set by a translational actuator. The opposite endpoint (B) is constrained to move along a vertical line $x=0$.

This system has one degree of freedom for each setting of the actuator; its state can be characterized by the angle $\delta$, which is a shape descriptor in this problem. Here $d_a=1$ and considering that $F$ can vary, $d_l=1$, hence eq. \eqref{eq:d:def} results in $d=2$. 
The line of action of the force $F$ is fixed. Note that the $l$ length of the spring is determined by the actuator length $x_l$ and the shape descriptor $\delta$ via
\begin{align}
\label{eq:dm:geomconst}
l\cos\delta-x_l=0.
\end{align}

Employing eq.~\eqref{eq:dm:geomconst}, length $l$ can be eliminated and the $\L$ total energy of the system follows:
\begin{align}
\label{eq:dm:lagrangian}
\L=\frac{1}{2}k\left(1-\frac{x_l}{\cos\delta}\right)^2+F\frac{x_l}{\cos\delta}\sin\delta,
\end{align}
where the first term is the elastic energy stored in the spring with unit unloaded length; the second term gives the potential of the external load $F$. Variation (i.e., the first derivative) w.r.t parameter $\delta$ delivers the equilibrium equation:
\begin{align}
\label{eq:dm:govern}
F-k\left(1-\frac{x_l}{\cos\delta}\right)\sin\delta=0.
\end{align}

\begin{figure}
    \centering
    \includegraphics[width=0.65\linewidth]{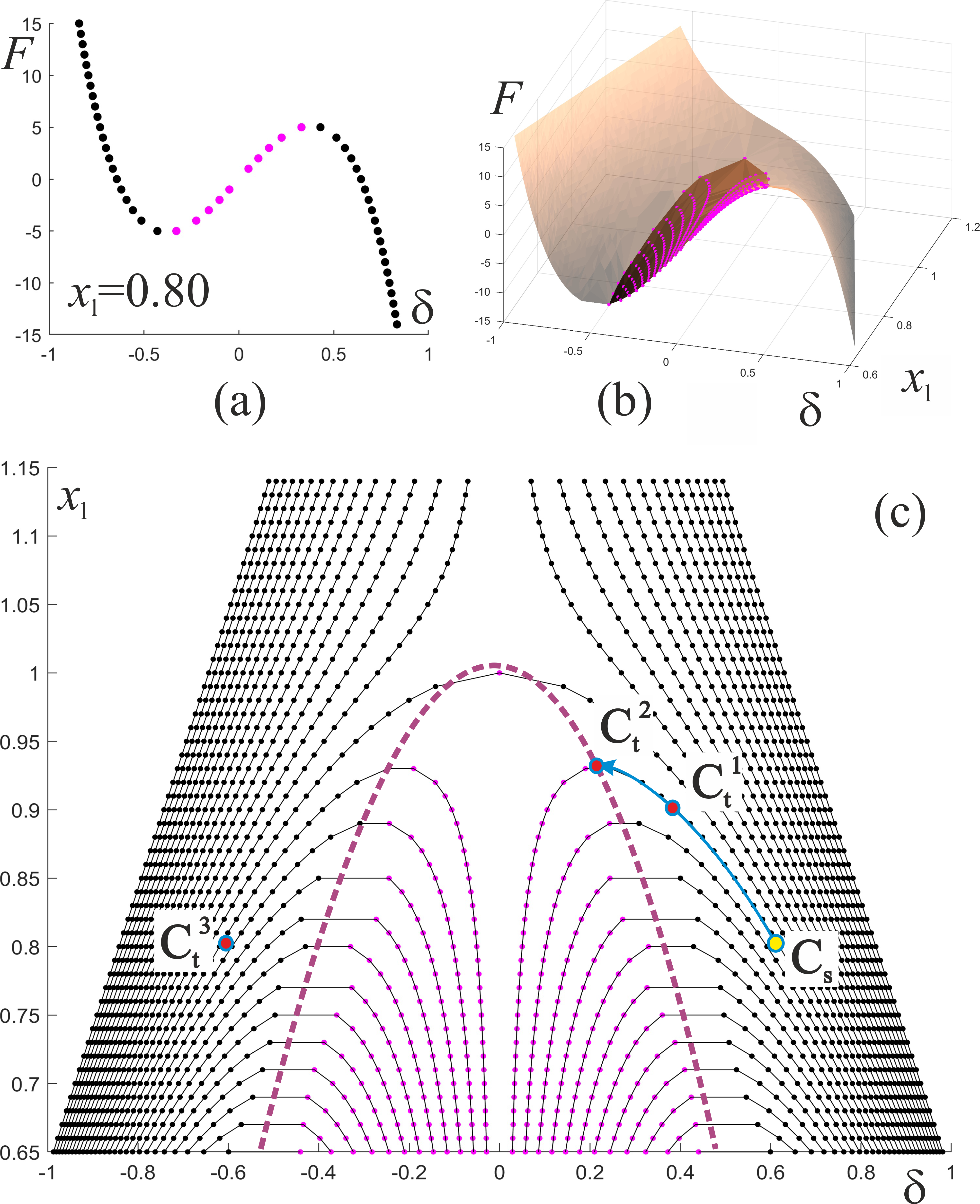}
    \caption{Illustration of adaptation tasks. (a) Equilibrium solutions of the problem in Figure~\ref{fig:dmodel} at a fixed setting of the actuator ($x_l=0.80$). Black and purple dots correspond to stable and unstable configurations, respectively. (b) The equilibrium set in the space spanned by $\delta$, $x_l$, and $F$. The region denoted by purple dots corresponds to unstable solutions. (c) The equilibrium set with level sets corresponding to constant $F$ values. The dashed purple curve, the stability boundary, separates stable and unstable solutions. In this problem, inverse kinematics from $C_s$ to $C_t^1$ and optimization until $C_t^2$ is possible, while path-planning to $C_t^3$ is impossible at a fixed level of the load.}
    \label{fig:dmodel:results}
\end{figure}

Stability of the solution is investigated with the help of the $D^2\L$ second variation of the Lagrangian, which is here the second derivative of $\L$, evaluated at solutions $\delta=\tilde\delta$ fulfilling eq.~\eqref{eq:dm:govern}:
\begin{align}
\label{eq:dm:secvar2}
D^2\L(\tilde\delta)=k\frac{x_l(x_l-\cos^3\tilde\delta)}{\cos^4\tilde\delta}.
\end{align}
A positive $D^2\L$ implies a stable solution. Eq.~\eqref{eq:dm:secvar2} shows that for $x_l>1$ every solution is stable, and we expect unstable solutions for $\delta$ close to 0 if $x_l<1$ holds. The equilibrium set is depicted in Figure~\ref{fig:dmodel:results}, where stable and unstable regions are marked by black and purple dots, respectively. Some adaptation tasks can be readily demonstrated. If one wants to drive the structure from the point $C_s:(\delta,x_l)=(0.62,0.80)$ to $C_t^1:(0.40,0.90)$ at a fixed load level $F=1$, then it is easily doable along the stable solutions connecting the two points.
This is an example of an inverse kinematics problem. If we aim to minimize $\delta$ with $C_s$ as initial state, then the optimal setting of the actuation corresponds to point $C_t^2$ at the \emph{stability boundary} (denoted by the purple dashed curve). 
In other words, the optimal state corresponds to $\delta=0.30$, which is achieved by actuation to $x_l=0.925$. Further actuation would lead the system to the unstable region and initiate dynamic snap-through. Finally, for this example, actuation to point $C_t^3:(-0.62,0.80)$ at a fixed level of the load is not possible, as no stable level-sets connect $C_s$ to $C_t$. This is an example of a path planning problem lacking a solution due to lack of connectivity. Nonetheless, if variation of the load was a viable option, one could design a path between the two points in such a way that the unstable submanifold is avoided.

\section{Application: a curved rod}  \label{sec:application}
\subsection{Problem formulation and equilibrium set}

In this section, we study an inextensible and unshearable Cosserat rod \citep{antman1995}, constrained to the plane. This model can be viewed as an ideal representation of a compliant arch structure created by deforming an initially straight structural element into a curved shape using incompatible boundary conditions - a technique commonly referred to in structural engineering as active bending \citep{lienhard2013active}. The rod is parameterized by its arc length $s$ and let $(.)_{,s}$ denote the derivative w.r.t. the arc length parameter. The computational domain is $\Omega:[0, L]$, where $L$ is the rod's length in the reference configuration. The local tangent direction is denoted by $\theta(s)$, and the physical coordinates of its points as $x(s),y(s)$. 

\begin{figure}[ht!]
    \centering
    \includegraphics[width=0.50\linewidth]{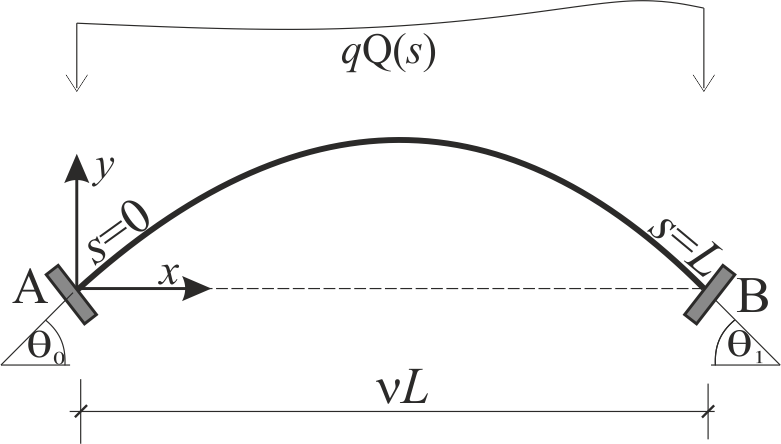}
    \caption{A continuous problem: a curved rod actuated at both of its endpoints $A$ and $B$ under a nonuniform load $qQ(s)$. The distance between the two endpoints of the rod is fixed.}
    \label{fig:cmodel}
\end{figure}

With $\kappa$ being the curvature, the deformed shape fulfills the following kinematic relations:  
\begin{align}
\label{eq:kappa}
\theta_{,s}&=\kappa,\\
\label{eq:x}
x_{,s}&=\cos\theta,\\
\label{eq:y}
y_{,s}&=\sin\theta.
\end{align}
The positions of the two endpoints at $s=0$ and $s=L$, denoted by $A$ and $B$, are assumed to be fixed at $(x(0),y(0))=(0,0)$, and $(x(L),y(L))=(\nu L,0)$, 
resulting in the following constraint equations:
\begin{align}
\label{eq:cons:horizontal}
\int_0^Lx_{,s}\d s&=\int_0^L\cos\theta(s)\d s=\nu L,\\
\label{eq:cons:vertical}
\int_0^Ly_{,s}\d s&=\int_0^L\sin\theta(s)\d s=0,
\end{align}
where $0<\nu<1$ is an a-priori fixed model parameter. 

The tangent angles of the rod at those terminal points are set by two rotary actuators as $\theta(0)=\theta_0$, $\theta(L)=\theta_1$. The rod has no intrinsic curvature; however, it forms a curved arch shape due to active bending induced by the kinematic constraints.

The elastic energy $U$ and the work of the external load $W$ read
\begin{align}
\label{eq:energy:elastic}
U&=\frac{1}{2}\int_0^LB\kappa(s)^2 \d s,\\
\label{eq:external_work}
W&=\int_0^L qQ(s)y(s)\d s,
\end{align}
where $B=EI$ denotes the bending stiffness, $q$ is a scalar parameter of the load intensity, and $Q(.):[0,L]\rightarrow\xR$ is the load distribution constrained to $\int_0^L Q(s)\d s=1$. 

The Lagrangian $\L$ of the system readily follows:
\begin{align}
\label{eq:total}
\L=U+W+\lambda\left(\int_0^Lx_{,s}\d s-\nu L\right)+\mu\left(\int_0^Ly_{,s}\d s\right),
\end{align}
where $\lambda$ and $\mu$ denote the Lagrange multipliers associated with the constraint equations (\ref{eq:cons:horizontal})-(\ref{eq:cons:vertical}). In physical terms, $\lambda$ and $\mu$ are the horizontal and vertical reaction forces at point $A$, respectively.

Before proceeding with further development, the following normalization is applied to nondimensionalize the problem. Let 
\begin{equation}
\lambda_{cr}:=\frac{4B\pi^2}{L^2}.
\end{equation}
denote the Euler critical load of the clamped-clamped straight elastica. Then $\theta\rightarrow\theta$ and apply $s\rightarrow sL^{-1},\quad x\rightarrow xL^{-1},\quad y\rightarrow yL^{-1},\quad \kappa\rightarrow \kappa L$ $B\rightarrow \frac{B}{\lambda_{cr}L^2},\quad Q\rightarrow QL,\quad q\rightarrow q\lambda_{cr}^{-1},\quad \lambda\rightarrow \lambda\lambda_{cr}^{-1},\quad \mu\rightarrow \mu\lambda_{cr}^{-1}$. Hence, from now on $L=1$ is the length of the rod and $\lambda=1$ corresponds to the Euler buckling load of the clamped-clamped rod. The non-dimensional Lagrangian $\L$ follows
\begin{eqnarray}
\label{eq:total:nondim}
\L=\int_0^1\left\{\frac{1}{2}\frac{1}{4\pi^2}\kappa^2 +qQy\right\}\d s+ \lambda\left(\int_0^1x_s \d s-\nu L\right)+\mu\left(\int_0^1y_s \d s\right).
\end{eqnarray}
Introducing $\rho(s):=\int_0^s Q(\zeta)\d\zeta$ and performing integration by parts for the second term, the functional reads
\begin{align}
\label{eq:total:nondim2}
\L=\int_0^1\left\{\frac{1}{2}\frac{1}{4\pi^2}\theta_{,s}^2 -q\rho\sin\theta\right\}\d s+
\lambda\left(\int_0^1\cos\theta \d s-\nu L\right)+\mu\left(\int_0^1\sin\theta \d s\right).
\end{align}
The $D \L$ first variation of the total energy for all admissible $\eta(s)$ variations, after integration by parts for the first term, is 
\begin{align}
\label{eq:first:variation}
D \L=\int_0^1\left\{-\frac{1}{4\pi^2}\theta_{,ss}-q\rho\cos\theta-\lambda\sin\theta+\mu\cos\theta\right\}\eta\d s.
\end{align}
This last expression readily delivers the equilibrium equation:
\begin{eqnarray}
\label{eq:total:equ}
\frac{1}{4\pi^2}\theta_{,ss}+q\rho\cos\theta+\lambda\sin\theta-\mu\cos\theta=0;
\end{eqnarray}
subject to non-dimensionalized form of the constraint equations (\ref{eq:cons:horizontal})-(\ref{eq:cons:vertical}) and the following boundary conditions:
\begin{eqnarray}
\label{eq:total:bcs}
\theta(0)=\theta_0, \qquad \theta(1)=\theta_1,
\end{eqnarray}
where $\theta_0$ and $\theta_1$ are fixed slopes at the ends of the curved rod.

The $D^2\L$ second variation of the total energy at an equilibrium solution $\theta(s)=\theta^*$ that fulfills (\ref{eq:total:equ}) is obtained:
\begin{eqnarray}
\label{eq:secvar:full}
D^2\L(\theta^*)=\int_0^1\left\{\frac{1}{4\pi^2}\eta_{,s}^2+q\rho\sin\theta^*\eta^2-\lambda\cos\theta^*\eta^2-\mu\sin\theta^*\eta^2\right\}\d s
\end{eqnarray}
for all admissible variations $\eta(s)$. As we investigate a planar problem, $\eta(s)$ represents planar perturbations of the shape. The stability of the arch hinges on the support reactions and the actual external load $q\rho(s)$. Compared with the classical solution of the elastica, the second variation of the Lagrangian in (\ref{eq:secvar:full}) suggests that stability at $\lambda>1$ cannot be excluded because of the presence of the external lateral loading. 

The equilibrium set might be parameterized internally by $(\theta_1,\theta_2,q)$, hence $d_a=2$ and $d_l=1$ resulting in $d=3$. For convenience, we introduce
$$
\Delta:=\frac{\theta_1+\theta_0}{2}, \Theta:=\frac{\theta_1-\theta_0}{2}
$$
as the mean and difference of the support angles. Using $(\Delta,\Theta,q)$ instead of $(\theta_0,\theta_1,q)$ was chosen because $\Delta$ represents the degree of asymmetry.

The equilibrium set of the arch can be constructed by computing the equilibrium configurations at all points of a sufficiently dense rectangular mesh in the $(\Theta,\Delta)$ parameter plane.

The constrained boundary value problem in (\ref{eq:total:equ}) has no known closed-form solutions, so we seek its stable solutions numerically. We discretize the structure by $n$ points and approximate the derivatives of $\theta$ at the grid points by the derivatives of Chebyshev polynomials \citep{Chebfun2014}. It is shown that the Chebyshev approximation converges exponentially with $n$ to the original function \citep{Trefethen2000}. The discretization results in $n$ equations that we solve using numerical continuation with the pde2path toolbox \citep{Uecker2021}. We monitor the sign of the eigenvalues of the Jacobian of the system to identify stable solutions and detect bifurcation points. Note that the rod is constrained to the $[xy]$ plane; hence, in this work, instabilities associated with out-of-plane deformations are not considered.






\subsection{Structural optimization of the curved rod}
\label{sec:optim}

A straightforward application of adaptivity is the optimization of the structural behavior in the case of time-dependent loads. Consider the curved rod with geometric parameters $L=1$, and $\nu=\sin(1)$.
A symmetric, uniform load distribution 
$$
Q_s(s)\equiv 1,
$$
and an asymmetrical, piecewise constant one
\begin{align}
Q_a(s)=
\begin{cases}
    2 & \text{if    } s\leq 1/2\\
    0 &\text{if    } s>1/2
\end{cases}
\end{align}
are compared.

Two specific scenarios will be investigated, with our goal being to demonstrate that the optimal values of  $\Theta$ and $\Delta$ strongly depend on the distribution of loads, the constraints, and the objective function. 

\subsubsection{Maximizing the limit load}
\label{sec:maxload}

The first scenario is maximizing the limit load of the arch under the requirement of stability, as well as a constraint on its curvature.
We begin by noting that the internal parameterization of the equilibrium set by $(\Theta,\Delta)$ is ambiguous, which is eliminated by the reduction of the equilibrium set.
Specifically, for each actuator setting and load distribution, we start from a stable elastica with the prescribed $\theta_0, \theta_1$ values and $\lambda=1e-8$, while $q=0$. Then we reach an initial point of the equilibrium set using parameter continuation in $\nu$. 
Finally, an equilibrium path is generated by parameter continuation using the load intensity as a parameter. The parameter continuation is interrupted if a critical point is reached where any of the following two feasibility constraints is violated:
\begin{enumerate}
\item [C1]: stability of the equilibrium 
\item [C2]: maximum curvature along the rod is below the threshold value of $10$.
\end{enumerate}
The critical point was successfully identified for each actuator setup, except for those ranges of the actuator angles, where C2 is violated by the unloaded configuration, hence the critical load is undefined. 

Finally, a threshold value of the load intensity $q_{\max}(\Theta,\Delta)$ is determined as the maximum of the load intensity parameter along the equilibrium path between the initial point with $q=0$ and the critical point. If the critical value is undefined, then $q_{\max}$ is also considered undefined.

\begin{figure}
    \centering
    \includegraphics[width=0.49\linewidth]{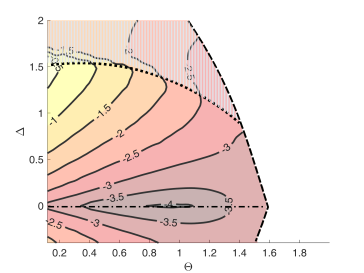}
    \includegraphics[width=0.49\linewidth]{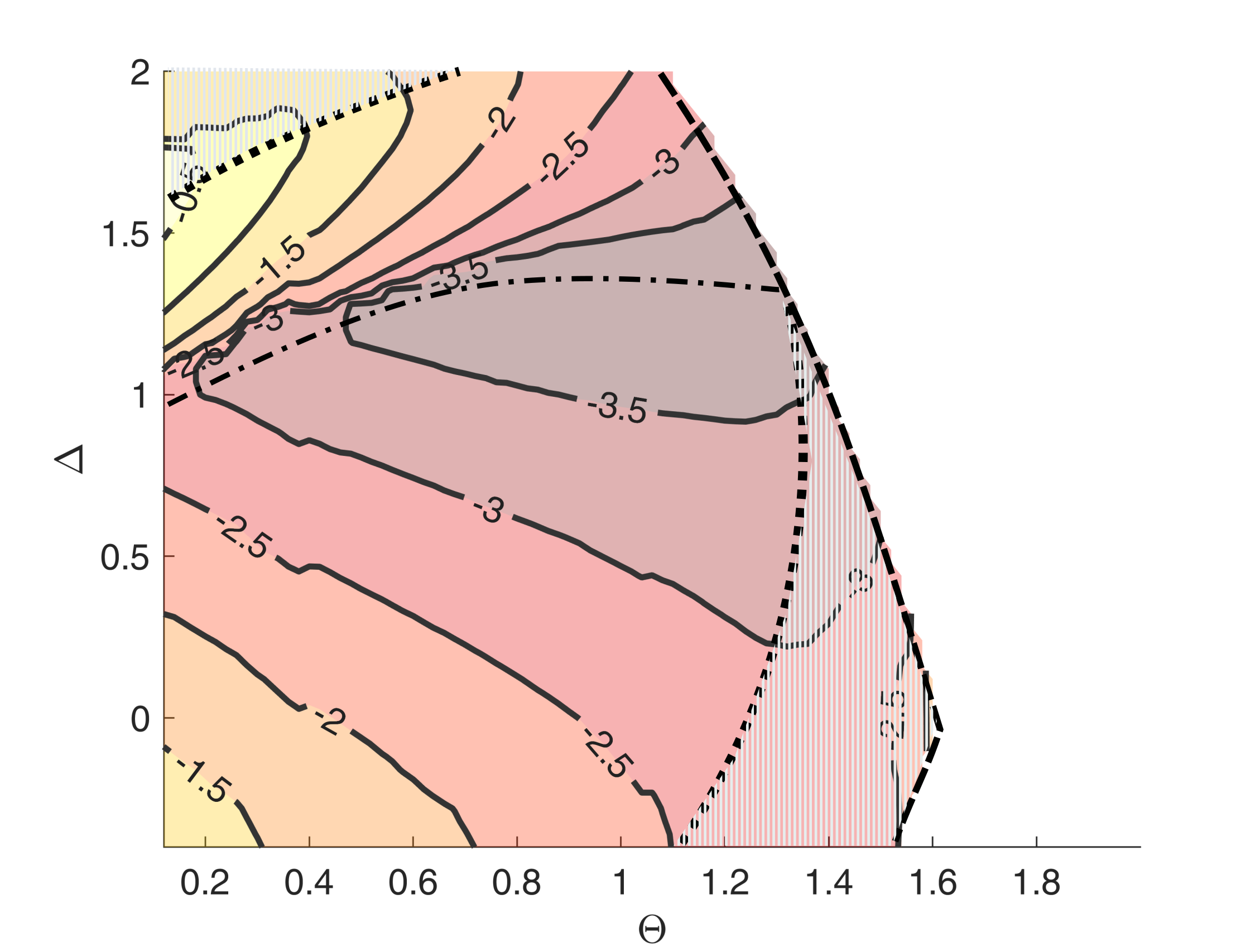}
    \\
(a) \hspace{0.4\textwidth} (b)
    \caption{ Contour plot of maximum load intensity $q_{\lim}(\Theta,\Delta)$ for uniform load distribution $Q_s$ (a) and asymmetrical load distribution $Q_a$ (b). Dashed line indicates those regions where condition C2 is not satisfied in the unloaded state; white overlay shows those regions where $q_{\lim}$ is determined by C2. Dash-dotted curves mark bifurcations of the equilibrium path resulting in non-smooth dependence of $q_{\lim}(\Theta,\Delta)$ on its variables.}
    \label{fig:limload}
\end{figure}

The results of this computation for the two load distributions $Q_s$ and $Q_a$ are illustrated by the contour maps of Fig.~\ref{fig:limload}(a,b).
The function $q_{\max}(\Theta,\Delta)$ is piecewise smooth in its variables with non-smooth or discontinuous behavior along those curves where a bifurcation of the equilibrium path occurs. Bifurcations have been identified by visual inspection of equilibrium curves and highlighted by dashed-dotted curves in the figures. 
 
 In most cases, the critical load (and thus $q_{\max}$) is determined by the stability requirement C1, however, for large values of $\Delta$ and of $\Theta$ the structure remains stable for all load intensity values, and thus $q_{\max}$ is determined by the curvature threshold C2. These regions have been highlighted by a white overlay bounded by dashed or dotted curves. The dashed boundaries of the plots mark those support angles beyond which the curvature threshold C2 is violated by the unloaded shape.

In the case of symmetrical load $Q_s$, the optimal actuator setting is also symmetrical ($\Delta=0$), with $\Theta\approx 0.9$, where the limit load intensity is $q_{\lim}(0.9,0)\approx-4.2$. Notably, the symmetrical configurations correspond to a bifurcation of the equilibrium set (corresponding to asymmetrical buckling), where the limit load is very sensitive to symmetry-breaking imperfections (such as $\Delta\neq 0$). For example, $q_{\lim}(0.9,0.04)\approx-3.9$. The high level of sensitivity has to be taken into account in most practical applications, which is a well-known result of elastic stability theory based on the theoretical foundations of mathematical catastrophe theory \citep{poston2014catastrophe,varkonyi2006symmetry}. Sensitivity has fundamental practical consequences to the design of thin shells \citep{arbocz2002future}, and thus it is not surprising that slender arches behave similarly. Robustness and systematic sensitivity analysis are beyond the scope of this paper, but they are important aspects of optimization by shape adaptation.

For the asymmetrical load distribution $Q_a$, we have $q_{\lim}(0.9,0)\approx-2.5$; however, the limit load can be increased considerably by optimizing the support angles. The optimal value of the limit load is $q_{\lim}(0.66,1.36)\approx-4.0$, which again lies on a bifurcation curve, and thus is also prone to extreme imperfection sensitivity.
Despite the sensitivity issues, this example demonstrates that the limit load can be increased significantly by tuning support angles in response to variations of the load distribution. 

\subsubsection{Minimizing the curvature} \label{sec:mincurve}

\begin{figure}
    \centering
    \includegraphics[width=0.49\linewidth]{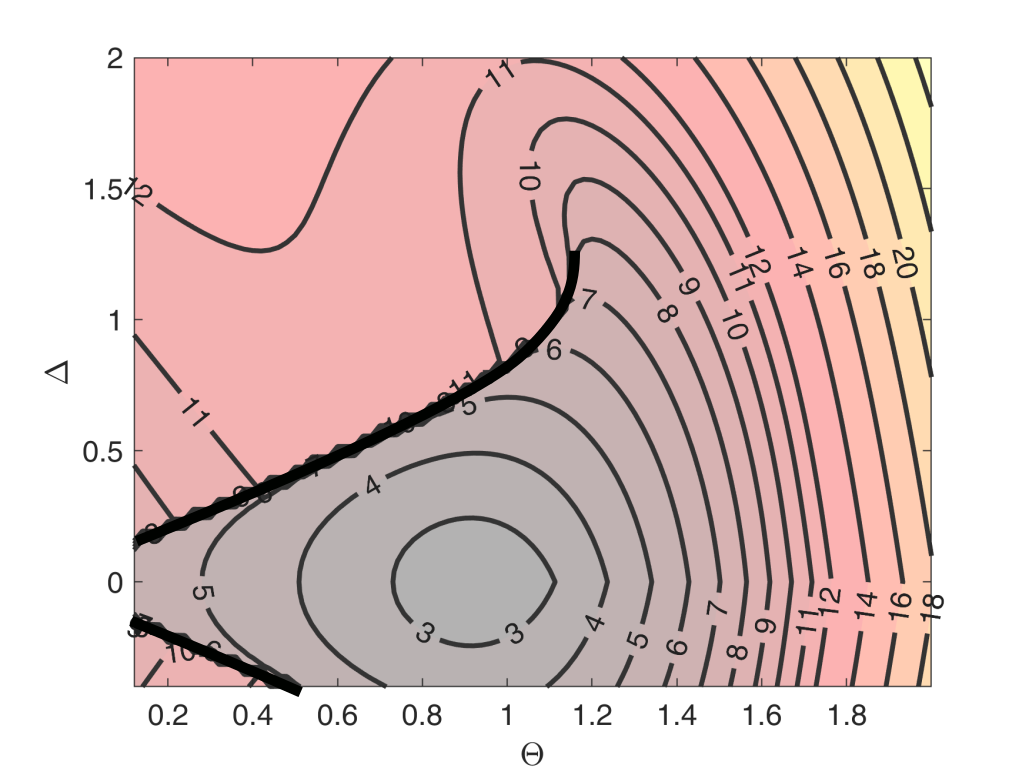}
\includegraphics[width=0.49\linewidth]{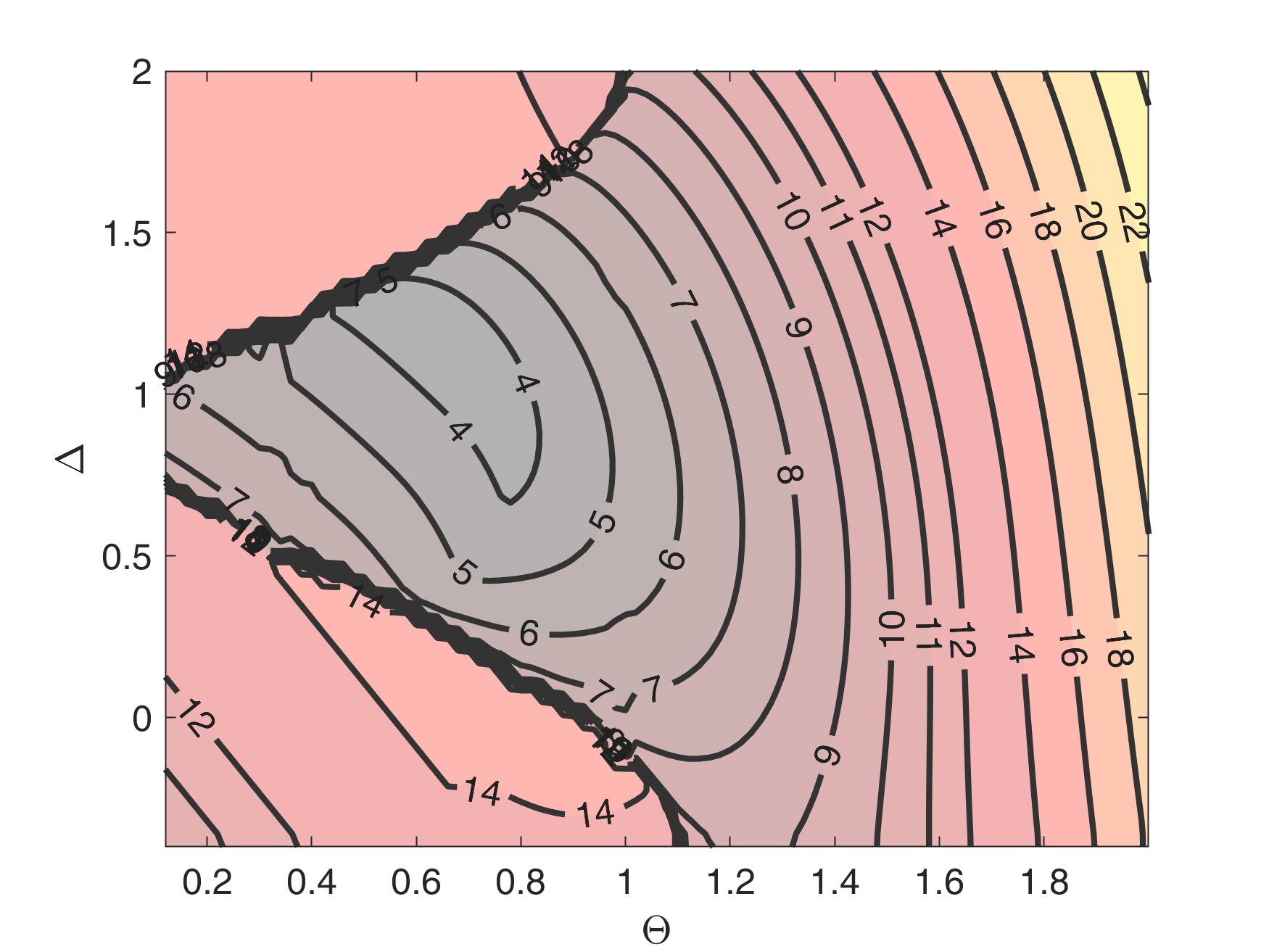}
\\
(a) \hspace{0.4\textwidth} (b)
    \caption{Contour plot of the largest curvature as a function of the support angles $\Theta$ and $\Delta$ for uniform load (a) and asymmetrical load (b) with intensity $q=2.50$. The thick black curves indicate  'barriers' in the projected equilibrium manifold.}
    \label{fig:maxcurv}
\end{figure}

The second scenario involves minimizing the (largest) curvature along the same arch for both load distributions at a fixed level of the load intensity $q=2.5$. This time, the structure is optimized for the objective value $O_1$ (as defined by eq. \eqref{eq:O1kappa}). The equilibrium set is parameterized internally  by $(\Theta,\Delta)$.  Non-uniqueness is eliminated by reduction, similarly to the previous scenario. In particular, starting from the same unloaded initial configuration as before, a parameter continuation algorithm is executed with $q$ as parameter, until a stable equilibrium with load intensity $q=2.5$ is reached for the first time. The objective function is evaluated numerically at this final configuration  
(Figure \ref{fig:maxcurv}). 

For the symmetric load, the smallest value of the objective function occurs at $(\Theta,\Delta) \approx (0.94,0)$.
In contrast, the optimal setting for asymmetrical load is $(\Theta,\Delta)\approx (0.64,1.16)$. For both settings, switching between the two load distributions induces more than a 3-fold increase in the curvature (Table \ref{tab:curvatures}). In addition, the asymmetric load combined with the symmetric actuator setup induces an intermediate segment of unstable equilibria along the equilibrium path. In practice, this situation means that the structure undergoes dynamic snap-through before the specified load level is reached. This example demonstrates the high potential of stress relaxation by shape morphing.
\begin{table}[]
    \centering
    \begin{tabular}{c c|c c} 
        $\Theta$&$\Delta$&Symmetric load & Asymmetric load  \\ \hline
        0.96 & 0 & 2.1& 7.6 \\
        0.64 & 1.16 & 11.7 & 3.35 
    \end{tabular}
    \caption{Maximum curvatures along the rod for two different load distributions, and two actuator settings at load level 2.5.}
    \label{tab:curvatures}
\end{table}

\subsection{A parametric inverse kinematics problem}
\label{sec:inversekinematics}
\begin{figure}
    \centering
    \includegraphics[width=0.49\linewidth]{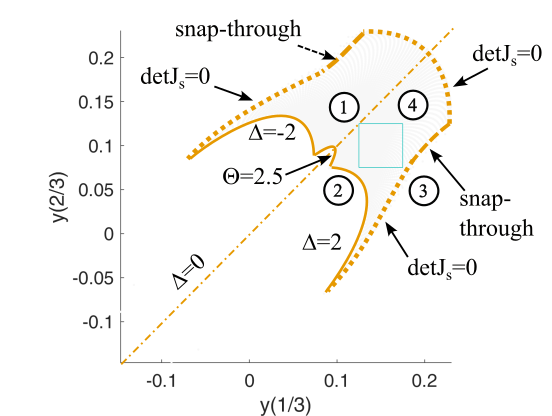} 
\includegraphics[width=0.49\linewidth]{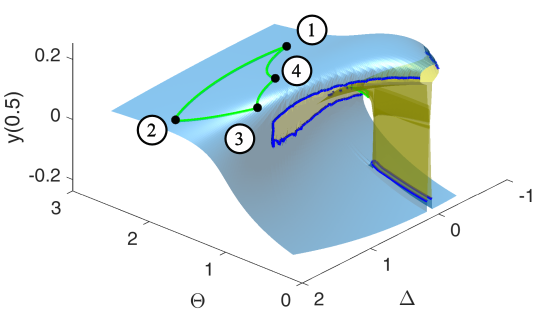}
\\
(a) \hspace{0.4\textwidth} (b)
 \\
\includegraphics[width=0.49\linewidth]{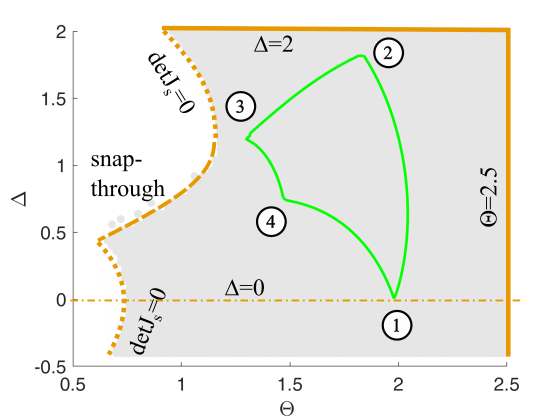}
\includegraphics[width=0.49\linewidth]{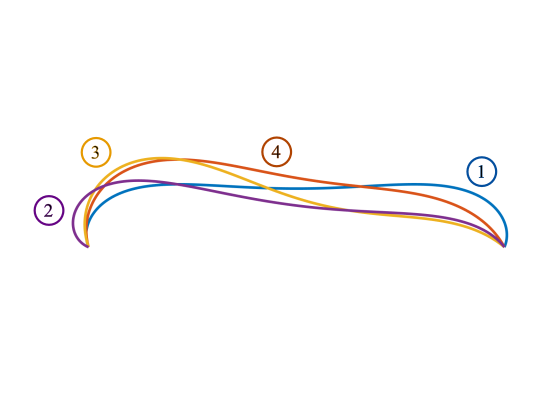}
\\
(c) \hspace{0.4\textwidth} (d)
    \caption{Tracking of a slowly changing target shape. (a): a shape trajectory, i.e., periodically changing prescribed values of $y(1/3)$, and $y(2/3)$ corresponding to a rectangle with vertices labelled by numbers 1,2,3,4.  (b): the solution of the corresponding inverse kinematics problem embedded into the equilibrium surface. (c): projection of the solution into the plane of actuation parameters (with the interval $-2\leq \Delta\leq -0.5$ omitted for better visibility).  (d): physical shapes of the elastic rod at four representative points of the solution curve. In (a,c), grey shading marks the range of reachable shapes available with the chosen actuator setup. This region is symmetric, with the symmetry axis corresponding to $y(1/3)=y(2/3)$, or equivalently to $\Delta=0$. The boundaries of the reachable region are marked by three types of curves: boundaries of unstable regions where snap-through occurs; singularities of the Jacobian $J_{s}$, and limits of the examined range of actuator parameters ($\Theta=2.5$; $\Delta=\pm 2$).}
    \label{fig:shapepath}
\end{figure}

This time, our goal is to track slowly varying target values of shape descriptor functions by tuning the two actuators of the arch. The case of a stationary, constant load with intensity $q=2.5$ is considered. We use the height of  points $s_1=1/3$, and $s_2=2/3$ along the curve as shape descriptors: $S_1=y(1/3)$, $S_2=y(2/3)$. The goal of shape adaptation is to track a one-parameter set of values forming a rectangle in $(S_1,S_2)$ plane, see Figure \ref{fig:shapepath}(a). The corresponding values of the time parameter $t$ are not important and are thus not specified. 

As a first step towards the solution, the pre-image of the target curve $(S_1(t), S_2(t))$ along the equilibrium set has been determined numerically by using linear interpolation. The pre-image forms a unique closed, piecewise smooth curve, which lies entirely in the stable region (Figure \ref{fig:shapepath}(c)). Panel (b) shows the projection of the same curves into $(\Theta,\Delta)$ plane. Some of the corresponding physical shapes are depicted in panel (d). 
These numerical results suggest that the parametric problem has a unique solution. 

As an additional goal, we also determined 
the reachable sets associated with the point $(S_1,S_2)=(0.15,0.15)$ as defined in Section \ref{sec:parametric}. 
First, the equilibrium set is reduced in the same way as in Sec. \ref{sec:mincurve} and internally parameterized by $(\Theta,\Delta)$. 
The reduced and parameterized equilibrium set is then segmented by numerical computation of the contour lines corresponding to \eqref{eq:detJ}. The previously found solution of the parametric problem is indeed contained in one single segment (Figure \ref{fig:shapepath}(b)) in line with our preliminary expectations. Finally, the reachable set is determined numerically by mapping this region numerically via $S(p)$ into the plane of shape descriptors (Figure \ref{fig:shapepath}(a)). 

\subsection{Path planning over the equilibrium set}

In the case of this task, we consider a stationary load function and seek a continuous path between two target points of the equilibrium set, 
while certain constraints must be satisfied. The two points may correspond to different actuator settings, or they can even be stable points for the same setting if the structure is multi-stable.  

\begin{figure}
    \centering
   \includegraphics[width=\linewidth]{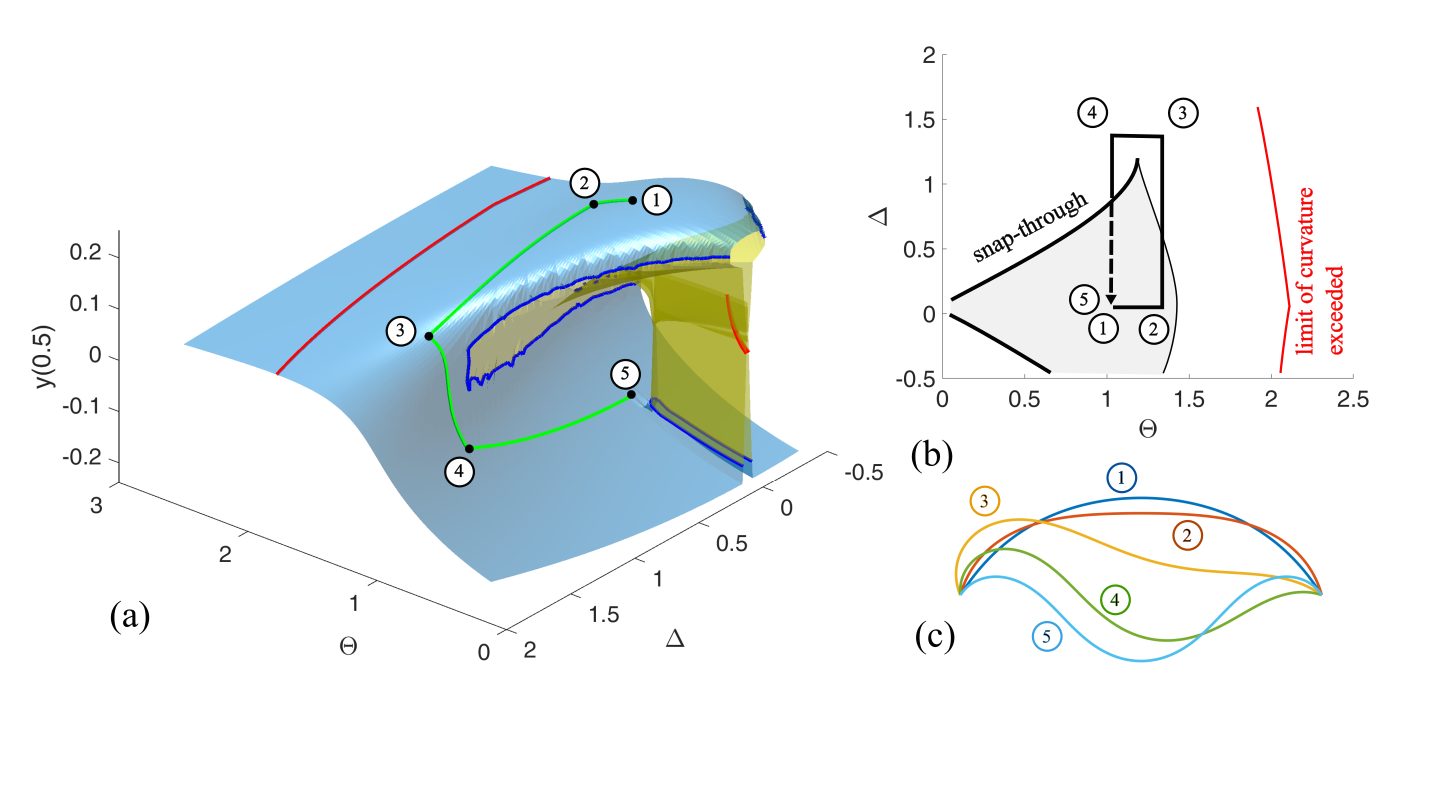} 
    \caption{(a): The equilibrium surface projected into $(\Theta,\Delta,y(0.5))$ space. Surface color marks stable (blue) and unstable (yellow) equilibria. A thick red curve indicates the threshold of curvature. A thick blue curve marks the boundary of unstable regions, whereas the green curve marks a path connecting points $1$ and $5$. (b): a 2D projection into $(\Theta,\Delta)$ plane where the projected path forms a rectangle. The region of multistability is highlighted by light grey shading. (c): physical shapes of the elastica, at representative points along the path}.
    \label{fig:path}
\end{figure}

We continue to use the elastic arch example with constant load distribution $Q_s$, and intensity $q=2.5$. The target points are two distinct stable equilibria for $\Theta=1$, $\Delta=0$, 
where the first one corresponds to an arch-like shape in which the load is balanced by bending and compression, and the second one to an inverted shape dominantly subject to tensile forces and bending. 
The equilibrium set is depicted in Fig. \ref{fig:path}(a), and the target points are labelled by numbers 1 and 5.
As before, a stability constraint and a threshold value of curvature $\kappa_{\max}=6$ are prescribed as feasibility constraints. The excluded regions of the equilibrium set are also depicted in the figure. 


A connecting path is constructed in the following steps. Visual inspection of the equilibrium set suggests that a piecewise linear path in $(\Theta,\Delta)$ plane defined by the points $(1,0)\rightarrow (1.3,0)\rightarrow (1.3,1.3)\rightarrow (1,1.3)\rightarrow (1,0)$ guides the structure through stable equilibria from point 1 to point 5. The corresponding curves along the equilibrium set can be found by interpolation, as shown in Fig. \ref{fig:path}(a). Notably, the structure is bistable for a range of actuator settings along the planned route; hence, interpolation delivers some fake solution segments, which lie on the wrong branch of the equilibrium set. These segments can be filtered out easily by cutting the solution curves at the target points. 

A more advanced approach to the problem would be to seek an optimal path between the endpoints using an appropriate objective \citep{sachse2021variational}. In order to use standard path panning algorithms, the unambiguous internal parameterization of the equilibrium set may be necessary. For this particular task, a combination of reduction (to remove unstable equilibria) and dissection into two parts (one containing the upper and the lower branch in the bistable region) is a natural choice. We do not elaborate further on this issue.

\section{Discussion and conclusion} \label{sec:conclusion}

Traditional design methodologies cope with future variation of the design parameters, including imperfections, variation of the external loads, and aging of the construction materials via \emph{robustness of the structure}, i.e., significant material reserves for unforeseen scenarios. In nature, on the other hand, adaptive responses have widely emerged during evolution. In other words, the time-independent approach of classical design methodologies might be extended towards solutions where changes in (load, material, etc.) parameters of the structure in service are handled as they emerge.

\citet{sobek2016ultra} argues that the idea of adaptive structures is a natural evolutionary step at the point where advanced structural form-finding methods uncover extremely lightweight designs lacking a relatively stable \emph{form-defining load case}, questioning the theoretical foundations of structural form-finding. In contrast to these arguments, adaptive structures are barely used in civil engineering at this time.

On the one hand, this vantage point challenges the half-probabilistic methods of structural design; on the other hand, it addresses the question of reliable control. Specifically, structural design in most cases assumes time-independent random variables as design parameters, and past data, collected in experiments and in service, determine the distributions of these random variables. Using the partial or safety factors, the risk of failure is controlled. In some cases (e.g., prestressed structures), this methodology is applied at different time instances of the structure's lifetime. In terms of probability, the application of time series and the explicit appearance of time in structural design is lacking. The idea of adaptive structures is not only accepting that some actions or material properties can vary unexpectedly during the service time, but it also investigates the possibility of in-built reactions to these occurrences. This paper is devoted to the realm of quasi-static shape adaptation, as it is the key to many practical applications. 

Compared to prior works, we highlight the challenges associated with morphing structures subject to large deformations, as well as extremely lightweight - and thus soft - structures.
The advantages of shape adaptation are revealed in these cases
promising essential material savings in construction. 

Specifically, the research presented in this paper shows that shape-adaptation with a finite number of actuators (a condition definitely needed for practical applications) might lead to a significant reduction in the internal stresses and curvatures, as well as an increase in load-bearing capacity without loss of stability. 

The computation of the equilibrium set not only allows for optimization or path planning but also lends itself to \emph{sensitivity analysis}. Traditionally, geometric imperfections are studied to characterize the robustness of the stability of the equilibrium solution. These investigations, for instance, are essential in the design of thin shells. In the case of adaptive structures, imperfections related to the sensors and actuators might be significant, so traditional approaches cannot be applied per se. Furthermore, optimal solutions might be in the close vicinity of bifurcations, as we have seen in the curved rod example of this paper. These findings hint toward a promising research question: opposed to the classical, local notion of stability, how to characterize the robustness of a stable solution in a non-local way? In terms of our paper, how to include the topology of the equilibrium set neighboring a (stable) solution in order to enhance the adaptation strategy.


\section*{Acknowledgments}
The support of grants TKP2021-NVA BME and K143175 by the National Research, Development and Innovation Office of Hungary is kindly acknowledged. We acknowledge the Digital Government Development and Project Management Ltd. for awarding us access to the Komondor HPC computing facility of Hungary. The authors declare that there are no competing interests relevant to this work.

\bibliographystyle{elsarticle-harv}
\bibliography{bibliography}

\end{document}